\documentclass[sigconf, nonacm]{acmart}

\setcopyright{none}
\renewcommand\footnotetextcopyrightpermission[1]{}
\acmConference[ACM SIGCOMM Computer Communication Review]{ACM SIGCOMM Computer Communication Review}
\acmYear{January 2023}

\usepackage{fancyhdr}
\fancypagestyle{plain}{%
   \fancyhf{} %
   \fancyfoot[L]{ACM SIGCOMM Computer Communication Review}%
   \fancyfoot[R]{Volume 48 Issue 1, January 2018}%
}
\fancypagestyle{firstpagestyle}{%
   \fancyhf{} %
   \fancyfoot[L]{ACM SIGCOMM Computer Communication Review}%
   \fancyfoot[R]{Volume 48 Issue 1, January 2018}%
}

\usepackage{tikz}
\usepackage{amsmath}
\usepackage{multicol}
\usepackage{etoolbox}
\usepackage{array} 
\usepackage{multirow}
\usepackage[para,flushleft]{threeparttable}
\usepackage{colortbl}
\usepackage{pifont}% http://ctan.org/pkg/pifont
\newcommand{\cmark}{\ding{51}}%
\newcommand{\xmark}{\ding{55}}%
\usepackage{graphicx}

\usepackage[frozencache,cachedir=minted-cache]{minted} % simple beautiful listing
\usepackage{booktabs}
\usepackage{enumitem}
\usepackage{times}

\definecolor{myc1}{rgb}{1,0.88,0.88}
\definecolor{myc2}{rgb}{0.2,0.88,0.2}
\definecolor{myc3}{rgb}{0.67,0.84,0.90}

\newcolumntype{x}[1]{>{\centering\let\newline\\\arraybackslash}p{#1}}
 
\begin{document}
%don't want date printed
\date{}

%\title{\Large \bf You must choose, but choose wisely:\\ The last crusade for the right RDMA-enabled data exchange algorithm}
\title{Efficient RDMA Communication Protocols} %\Large \bf 

\author{Konstantin Taranov}
\affiliation{
	\institution{ETH Zurich}
         \country{Switzerland}
}
\email{ktaranov@inf.ethz.ch}

\author{Fabian Fischer}
\affiliation{
	\institution{ETH Zurich}
          \country{Switzerland}
}
\email{fischerf@ethz.ch}

\author{Torsten Hoefler}
\affiliation{
	\institution{ETH Zurich}
          \country{Switzerland}
}
\email{htor@inf.ethz.ch}

\begin{abstract}
%More and more systems adopt RDMA networking to build highly scalable and efficient data processing, high-performance computing, and data query systems.
%
Developers of networked systems often work with low-level RDMA libraries to tailor network modules to take full advantage of offload capabilities offered by RDMA-capable network controllers.
Because of the huge design space of networked data access protocols and variability in capabilities of RDMA infrastructure, developers tend to reinvent and reimplement common data exchange protocols, wasting months of development yet missing various performance and system capabilities.
In this work, we summarise and categorize RDMA data exchange protocols and elaborate on what features they can offer to networked systems and what implications they have on their memory and network management. %Our analysis provides design guidelines for building best-fitting communication modules for networked systems  given their design requirements and capabilities of used RDMA infrastructure. % and used network controllers.

\end{abstract}

\begin{CCSXML}
<ccs2012>
   <concept>
       <concept_id>10002951.10002952.10002971.10003450</concept_id>
       <concept_desc>Information systems~Data access methods</concept_desc>
       <concept_significance>300</concept_significance>
       </concept>
   <concept>
       <concept_id>10003033.10003034</concept_id>
       <concept_desc>Networks~Network architectures</concept_desc>
       <concept_significance>300</concept_significance>
       </concept>
 </ccs2012>
\end{CCSXML}

\ccsdesc[300]{Information systems~Data access methods}
\ccsdesc[300]{Networks~Network architectures}

\keywords{RDMA, Networked Systems, System design}
 
\maketitle

%%% TODO. 
% I have top down approach where I take systems and then say what communication they required.
% NOw I need bottom up. I should think what systems I can build from the algorithms. 
% Or I can think about one system type and how we design it depending on features. 

% Problems
% Developers do not know all possible protocols and implement most popular/most simple one. 
% It is not clear which RDMA exchange is used and why. 
% Papers focus mostly on performance of low-level features or of overall systems. We aim to show performance of message exchange protocols. 
% Plus now we have diversity of RDMA protocols that have different capabilities and features. We provide hardware requirements to protocols. 
% We provide guidelines and tricks to improve protocols. We show 5 common tricks and 3 new protocols that are not listed by other papers. 
% We categorise all protocols. 

\section{Introduction}

RDMA is a prominent network mechanism that accelerates modern distributed systems~\cite{farm,naos,ramcloud,infiniswap,derecho}. 
It reduces CPU load and provides additional data access capabilities that are performed over a network. 
Despite the fact that RDMA is exploited by numerous systems, the community lacks a definitive study on 
algorithms for enabling communication between endpoints. 
Existing studies~\cite{anujRDMA,herd,Dragojevic2017RDMART} predominantly focus on the performance characteristics of RDMA systems and often omit design artifacts and features of employed communication protocols.  
The lack of a proper analysis of the capabilities of end-to-end protocols leads developers to blindly adopt RDMA-based data exchange algorithms without considering requirements imposed by the designed system, thereby often compromising the capabilities of the network modules of RDMA-enabled systems.

In this work, we provide a comprehensive list of possible RDMA-based data exchange protocols and study them according to the data and network management requirements posed by modern RDMA-enabled systems. 
We show there is no one-size-fits-all best approach for data exchange over RDMA networks and systems should always choose the algorithms according to their design requirements and the capabilities of utilized RDMA infrastructures.
For that, we analyzed numerous RDMA-enabled systems and found their requirements for data exchange channels (see Section~\ref{sec:systems}). 
%Our analysis of RDMA-enabled channels, in addition, lists typical optimizations used by developers, such as the use of specialized hardware capabilities of modern network controllers, to make their protocol faster and easier to implement. 
%Our study is not limited to InfiniBand-based transport protocols~\cite{infiniband} but also covers new RDMA-capable transports such as 1RMA~\cite{1rma} and EFA~\cite{efa}. 

We also introduce novel data exchange algorithms that, to the best of our knowledge, have not been studied or even proposed before. First, we explain how RDMA messaging via a circular buffer filled with RDMA writes can be implemented with a single RDMA write per message without the need for zeroing received records (see Section~\ref{sec:rings}). Previously a sender was required to issue at least two RDMA requests per message or a receiver to clear processed records. 
Second, we introduce a novel approach of sharing a single circular buffer between several remote senders, reducing memory usage for the N:1 communication (see Section~\ref{sec:sharedrings}). 
Finally, we propose a READ-centric circular buffer that allows the sender to "passively" broadcast messages to remote endpoints without the need to access the local RDMA device (see Section~\ref{sec:readrings}).

%Given our study, we designed a tool helping developers to choose data exchange channels according to system requirements and network capabilities (\textit{Contribution \#3}). 
%Besides facilitating correct design of new systems, the tool can be used to identify systems whose communication channels contradict the core requirements of primary workloads. 
%In particular, we argue in Section~\ref{sec:newsystems} that existing RDMA-enabled swapping systems~\cite{infiniswap,far} exploit memory-exhausting data channels, thereby conflicting with the objective to improve memory utilization in clusters. Thus, we propose data channels that significantly reduce the memory usage of swapping daemons without compromising their performance (see~Section~\ref{sec:newsystemsbench}). 

%that the developers of Infiniswap~\cite{infiniswap} mischose an RDMA exchange protocol for their system, in our opinion.  The system was designed with performance as the first resident and that the cluster can waste memory for that. In our opinion, a paging system should also provide efficient use of memory resources. We introduce FiniteSwap, a swapping system that significantly reduces the memory usage of infiniswap without compromising its performance. 

%\textbf{What this work is not about}. 
This work does not aim to find one "best" data exchange protocol that could be adopted by all networked systems. Instead, we show that this protocol does not exist and systems should choose a protocol depending on their needs and requirements. 
While many researchers (e.g.,~\cite{anujRDMA,herd,Dragojevic2017RDMART}) focus on tuning performance of protocols by finding optimal batch and work request sizes for certain RDMA infrastructures, we focus on \emph{fundamental features of protocols}, such as the number of round trips and support of interrupts, that are universal across RDMA hardware providers and will not change with the next generation of RDMA products (see Section~\ref{sec:rdmafuture}).

%we focus on \emph{fundamental performance indicators} that apply to all existing and future RDMA products, such as the number of round trips and the number of RDMA requests, and 
%We do not aim to introduce the best fine-tuned high-performance implementation of each protocol by finding optimal batch and work request sizes, because such studies has been already conducted by and the performance of RDMA requests may significantly differ across existing RDMA infrastructures.
%Thus, some algorithms can perform slightly better on one hardware than the others. Instead, 

\section{RDMA networking}
Remote Direct Memory Access (RDMA) is a network mechanism that empowers applications to access memory of remote processes without the involvement of their CPUs. 
RDMA-capable network controllers (RNICs) perform memory accesses using dedicated DMA controllers, allowing specialized network transport protocols to write incoming packet payloads directly to the local memory and fetch outgoing payloads from the memory, bypassing the CPU.

RDMA network protocols use an asynchronous programming model, where applications submit non-blocking communication \emph{work requests} to local RNICs, that use DMA to perform them without CPU involvement, and then asynchronously fetch corresponding completion events. The application can submit many communication requests in parallel without waiting for their completion, thereby achieving higher utilization of networking hardware.
As a result, the CPU load is much lower with RDMA networks than with classical networks employing the blocking POSIX API. 

%\subsection{RDMA protocols}
\label{sec:rdma}
In this work, we primarily focus on the IB verbs library~\cite{rdmacore}, the most popular standard for RDMA programming, which defines RDMA requests and RDMA hardware capabilities that can be supported and implemented by RDMA infrastructure. 

\textbf{RDMA requests.} RDMA transport protocols can fundamentally offer  SEND, WRITE, READ, and ATOMIC network requests to access remote memory and a local RECEIVE request to control destination memory of incoming SEND requests. SEND packets do not contain information about remote destination addresses, therefore, pre-posted RECEIVE requests are required to perform data reception. Unlike SEND, other requests are equipped with the destination address, allowing them to be executed without the control of the remote application (such access semantics is often called "one-sided access"). WRITE and READ requests allow writing content of local buffers to remote buffers and reading the content of remote buffers to local buffers, respectively. ATOMIC requests are requests that allow performing compare-and-swap and fetch-and-add over the network.%.\footnote{The IB verbs defines atomic requests of 4, 8, and 16 bytes.}. %, but most hardware providers only implement 8 byte atomics.

The IB library also defines WRITE\_WITH\_IMM requests that extend silent WRITE requests to generate completion events at the receiver, but at the cost of involving the receiver's CPU, which submits a RECEIVE request. The generated completion event contains the length of the WRITE request and an integer, called immediate data (IMM), specified by the sender.

\textbf{RDMA transports.} 
All discussed RDMA requests are supported by InfiniBand-based protocols: RoCE and InfiniBand~\cite{infiniband}. These protocols traditionally
%\footnote{Latest InfiniBand networks can be configured to use relaxed packet ordering to make use of adaptive routing~\cite{mellanoxOOO,mellanoxOOO2}} 
ensure in-order packet delivery, though each message can consist of multiple packets. For the in-order delivery, static routing of packets and full message re-transmission in case of packet drops are employed. 
As a result, InfiniBand-based protocols may suffer from head-of-line blocking and network congestion~\cite{rdma-congestion}. 

The EFA~\cite{efa} protocol alleviates these problems by adopting adaptive routing and allowing messages to be delivered out-of-order. With adaptive routing, packets may take up different paths in a network. To address the head-of-line blocking, each message of the EFA protocol consists of a single packet. 
However, the existing version of the protocol only supports SEND and READ requests. 

Similarly, 1RMA~\cite{1rma}  limits messages to one packet to enable out-of-order packet processing and reduce congestion. Additionally, 1RMA does not support SEND and WRITE requests as such packets with large payloads can overload the target RNIC or even cause side-effects such as unacknowledged memory writes (i.e., the target executes memory write but the initiator thinks that it failed). % if they are not expected
Fundamentally, the protocol only supports READ requests, however,
%which are always expected at the receiver side, as it is the initiator of the READ request. Nonetheless, 
the protocol offers emulation of WRITE requests that are implemented as requests to READ, therefore, requiring two full round trips.  

RDMA has a long history in the high-performance computing field,
where it was exposed via lower-level interfaces such as DMAPP or uGNI in Cray interconnects~\cite{slingshot}, libfabric~\cite{libfabric},  UCX~\cite{ucx}, or Portals~4~\cite{portals4}. 
These interfaces aim at run-to-completion applications like MPI~\cite{mpi3rma} and are only offered by specialized clusters (e.g., Cray supercomputers) while we aim to support client-server applications that can be deployed on commodity RNICs and in public clouds. 

\textbf{RNIC capabilities.} 
RNICs implement various mechanisms that are used by applications to lower latency, increase throughput, and improve memory management.
\textit{Scatter-gather list} allows applications to specify source and receive buffers as a list of non-contiguous buffers. 
\textit{Shared receive} allows applications to share RECEIVE requests between multiple connections.
\textit{Device memory} allows applications to allocate buffers in the memory of RNICs for networking.
\textit{Inline requests} allow senders to inline data to send work requests and receivers to inline incoming payloads to receive completions, removing one DMA transaction.
\textit{On-demand paging} empowers RNICs to dynamically fetch virtual to physical memory translation entries, removing the need to explicitly register the memory.

%	\item  \textit{Physical memory}: RNICs empower the endpoints to use physical addresses in RDMA requests, thereby removing the need to perform memory translation.

%All these capabilities have been extensively studied by other works~\cite{anujRDMA,herd,corm,Dragojevic2017RDMART,zero-copy-sges}, but they focus on performance optimizations for RDMA communication. 
%In this work, we aim to study fundamental differences between data exchange protocols, that could be enabled with these capabilities.

\begin{table}[t]
\centering
\setlength\tabcolsep{3.5pt}
\resizebox{\linewidth}{!}{%
\begin{threeparttable}
 \begin{tabular}{l|c|c|c}
Protocol & Requests\tnote{a} & Message delivery & Messages \\
\hline
\multirow{2}{2cm}{InfiniBand and RoCE~\cite{infiniband}}  & \multirow{2}{2cm}{Send, Write, Read, Atomic} & In-order\tnote{b} & multi-packet \\
 & & \\
          \hline
EFA~\cite{efa} 
         & Send, Read & Out-of-order & single-packet \\  \hline
1RMA~\cite{1rma}  
         & Read, Write\tnote{c} & Out-of-order & single-packet \\ 
\bottomrule
\end{tabular}
\begin{tablenotes}
\item[a] Not all requests must be implemented by RNIC providers.
\item[b] Some RNICs offer out-of-order delivery~\cite{mellanoxOOO,mellanoxOOO2}.
\item[c] A write is implemented as a request to read.
\end{tablenotes}
\end{threeparttable}
}
\caption{Features of RDMA-capable protocols.}
\vspace{-0.75cm}
\label{tab:mult-ways}
\end{table}

\textbf{Data ordering.}
Three factors can affect the delivery order of transmitted bytes: \emph{message ordering, packet ordering of a single message, and DMA ordering.}
In-order message delivery (e.g., InfiniBand) guarantees the delivery of messages in the order they have been issued by the sender. Some protocols (e.g., EFA and 1RMA) may reorder messages due to adaptive network routing. If a message consists of multiple packets, some protocols (e.g., InfiniBand with adaptive routing) can reorder packets of the message, resulting in out-of-order data placement at the receiver. EFA and 1RMA do not support multi-packet messages at all. 
However, even if packets are delivered in order, it is not guaranteed that DMA operations are performed in the same order.
%\footnote{Developers can check whether its RDMA connection supports in-order data delivery with the \texttt{ibv\_query\_qp\_data\_in\_order} call.}. 
Applications can intentionally relax DMA ordering to memory regions to improve the performance of RDMA requests to them. 
In our analysis, \emph{we say an RDMA protocol offers in-order byte delivery of a message if it ensures in-order delivery of its packets and in-order DMA operations.}

\section{Communication in RDMA systems}\label{sec:comprot}
We focus on data access and data exchange protocols used by RDMA-enabled systems.
We look at each data exchange as a \emph{uni-directional data channel} that allows a sender to communicate data to a receiver. 
Essentially, all data exchange protocols need at least two uni-directional data channels to acknowledge the  data reception for reliable data delivery.
Exchange protocols often use network acknowledgments of reliable transports to constitute an implicit uni-directional data channel. 
Alternatively, systems explicitly send data via another uni-directional channel to acknowledge message reception. 
Some high-level protocols can piggyback acknowledgments with data sent via a multi-purpose uni-directional communication channel. For example, RPC replies can be used as an indicator of the RPC request reception. 
An alternative approach is to build a dedicated channel for only acknowledging the data reception. 
%Often these channels can batch acknowledgments to reduce network load.
%We only focus on uni-directional data channels as they are a basic fundamental building block that can be used to build any data exchange protocol.

\emph{Overall, applications can build various data exchange protocols by combining explicit and implicit uni-directional data channels, making them fundamental building blocks of data communicators.}

%\subsection{Protocol features requested by systems}
\label{sec:systems}
Modern RDMA-enabled systems pose various requirements to network modules and underlying RDMA exchange protocols. 
To identify these requirements we analyzed numerous RDMA-enabled systems (see "Related Work" in Table~\ref{tab:protocols}) and reviewed fundamental features of their data exchange channels. We list these features below and elaborate on their importance in the system design.

\textbf{Performance. }  
Even though existing systems often argue about their performance in terms of latency and throughput, the performance of data exchange protocols is not portable across different RDMA networks.
We propose to measure the performance of protocols in metrics that are independent of RDMA infrastructure and portable across RDMA networks. Instead of the latency, we count \emph{the number of half-round-trips (HRTs)} required to send data via a data channel. Instead of the throughput, we count \emph{the number of issued RDMA network requests}. Thus, the latency of the protocol is proportional to the number of HRTs and the throughput is inversely proportional to the number of issued RDMA requests. For example, SEND and WRITE requests fundamentally perform alike as they require one request and one HRT to deliver data, even though WRITEs are slightly faster than SENDs in practice~\cite{herd}. 

\textbf{Blocking receiver.} RDMA requests offer two ways of notifying the receiver about incoming data: the receiver can be notified with a completion event or it can poll/read a special value from memory indicating message reception. While memory polling is reported to provide a better performance, it wastes a lot of CPU cycles on actively loading data from the memory. The completion event, on the other hand, can be received in blocking mode via interrupts.
As a result, the receiver can block at the completion events waiting for an interrupt, saving CPU cycles. For example, the NVMe-oF kernel module~\cite{linux-nvme} blocks on receive completions to reduce the CPU load. 

\textbf{True zero-copy.} Data copies waste a lot of CPU cycles and pollute CPU caches, encouraging systems to avoid them. Formally, if an initiator knows the source address of data and a protocol allows the sender to send this data without copies then we call the protocol \emph{zero-copy on send}. If a communication initiator knows the destination address of data and a protocol allows the receiver to get data to that address without additional copies then the protocol is \emph{zero-copy on receive}. 
For example, SEND-based protocols are not zero-copy on receive, as the sender has no control over the destination address at the receiver.
Thus, if we want to build a sender which must deliver data to a specific NUMA region depending on a hash of a message, we cannot do it with a single SEND-based protocol, and the receiver needs to copy data from pre-posted receive buffers to its final destination.
Note that the communication initiator can be the sender or the receiver depending on the protocol (e.g., READ initiators are data receivers, and their passive targets are data senders). 

\textbf{Variable size messages.} Not all protocols can efficiently manage memory for the reception of variable size messages and can only receive messages of one predefined size. 
Such protocols force applications to pre-allocate sufficiently large receive buffers fitting all possible message sizes, thereby increasing the total memory usage of applications.
For example, the main limitation of a SEND request is that it fully uses the pre-posted buffer at the receiver, even if incoming data was smaller than the receive buffer, causing significant memory fragmentation for variable messages. 
%Applications are forced to submit large receive buffers when they works with sizes.  

\textbf{Memory need for N:1 communications. } 
Modern systems may include several communication nodes. Of course, any two communicating nodes of a cluster can establish a point-to-point datapath, but it leads to significant memory usage that grows with the number of remote nodes. For example, if one receiver needs to receive messages from N nodes using point-to-point datapaths it will require managing receive buffers for each sender independently. 
For the N:1 communication, we later present protocols which enable efficient memory sharing of receive buffers between independent senders. 
For the 1:N communication, which multicasts messages to N nodes, all studied protocols can efficiently share their data source.

\textbf{Passive networking}. 
It describes whether a networked system can manage a communication channel without explicitly working with the RNIC (i.e., without submitting verb requests and polling queues). It means that systems can send and receive data using only load and store CPU instructions, thereby removing contention from the RNIC's queues.
Formally,  \emph{a sender supports passive networking} if it does not submit any work requests to the RNIC (see "\# Requests Send" in Table~\ref{tab:protocols}). \emph{A receiver supports passive networking} if it does not submit any work requests to the RNIC and also does not support blocking receive (see "\#~Requests Recv" and "Blocking Recv" in Table~\ref{tab:protocols}), which entails polling completion queues.

\begin{table*}[t]
\setlength\tabcolsep{2.0pt}
\centering
\resizebox{0.97\linewidth}{!}{%
\begin{threeparttable}[t]
\begin{tabular}{l|r|c|c|c|c|c|c|c|c|c|c|c|c|}
Protocol & Protocol Type & Latency & \multicolumn{2}{c|}{ \# Requests} & Blocking   & \multicolumn{2}{c|}{Zero-copy} & Variable &  \multicolumn{2}{c|}{Memory} &   Related &  \multicolumn{2}{c|}{RDMA transport requirements} \\ 
                 Family &   & \# HRT & Send & Recv   &   Recv & Send & Recv & Size & 1:$N$ & $N$:1 &  work & Requests & In-order   \\ \midrule \midrule
\multirow{3}{*}{Send/Recv} & Normal          & 1   & 1 & 0 & \cmark & \cmark & \xmark &\xmark & $O(1)$ & $O(N)$ & \cite{pilaf,infiniswap,anujRDMA,darpc} & Send & No   \\  
                           & Shared receive       & 1   & 1 & 0 & \cmark & \cmark & \xmark &\xmark & $O(1)$ & $O(1)$ &  \cite{ramcloud,hermes,fasst,tree-rdma,ccKVS}  & Send & No \\
                           & Bufferless      & 1   & 1 & 0 & \cmark & \textbf{N/A} & \textbf{N/A} & \textbf{N/A} & $O(1)$ & $O(1)$ &   \cite{anujRDMA,naos}  & Send & No  \\ \hline
\multirow{4}{*}{Write slot} & Detached bell    & 1   & 2 & 0 &  (\cmark/\xmark)\tnote{a} & \cmark & \cmark &\xmark & $O(1)$ & $O(N)$& \cite{l5} & Write & Messages   \\ 
                           & Inlined bell      & 1   & 1 & 0 & \xmark & \cmark\tnote{b} & \cmark\tnote{b} &\xmark & $O(1)$ & $O(N)$&  \cite{wukong,scaleRPC,deepRDMA,l5}  & Write & Bytes \\ 
                           & IMM capabilities        & 1   & 1 & 0 & \cmark & \cmark & \cmark &\xmark & $O(1)$ & $O(N)$& \cite{octopus,gam,l5,storm} & Write & No \\ 
                           & + reserve slot  & 3   &$\geq2$\tnote{a} & 1 & (\cmark/\xmark)\tnote{a}& \cmark & \cmark &\xmark & $O(1)$ & $O(1)$&   \cite{tailwind}  & Write & (No/Yes)\tnote{a}   \\  \hline
\multirow{4}{*}{Ring buffer} & Detached bell & 1   & 2 & 0 & \xmark &\cmark\tnote{b} & \xmark &\cmark & $O(1)$ & $O(N)$ &   \cite{dare}  & Write & Messages  \\ 
                           & IMM capabilities        & 1   & 1 & 0 & \cmark & \cmark & \xmark &\cmark & $O(1)$ & $O(N)$& \cite{naos}  & Write & No  \\ 
                           & Inlined bell  w/ zeroing & 1&1& 0 & \xmark & \cmark\tnote{b} & \xmark &\cmark & $O(1)$ & $O(N)$ & \cite{farm,drtmh,apus} & Write & Bytes  \\ 
                           & Inlined bell  w/o zeroing  & 1&1& 0 & \xmark & \cmark\tnote{b} & \xmark &\cmark & $O(1)$ & $O(N)$& this work & Write & Messages, Bytes \\ \hline
\multirow{2}{1.7cm}{Shared Ring buffer} & IMM capabilities   & 3&$\geq2$& 0 & \cmark &  \cmark & \xmark &\cmark & $O(1)$ & $O(1)$&  this work & Write, Atomic & No\\  
                           & Inlined bell  w/ zeroing & 3&$\geq2$& 0 & \xmark &  \cmark\tnote{b} & \xmark &\cmark & $O(1)$ & $O(1)$ &  this work & Write, Atomic  & Bytes\\ \hline
\multirow{4}{*}{Read slot} & Detached bell    & $\geq4$\tnote{c} &0& $\geq2$\tnote{c} & \xmark &  \cmark & \cmark &\cmark & $O(1)$ & $O(1)$& \cite{xstore,wukong,tree-rdma} & Read & No \\ 
                             & Inlined bell       & $\geq2$\tnote{c} &0& $\geq1$\tnote{c} & \xmark &  \cmark & \cmark &\xmark & $O(1)$ & $O(1)$& \cite{pilaf,infiniswap,far,derecho} & Read & No\\ 
                             & Notify   receiver   & 3& 1 & 1& (\cmark/\xmark)\tnote{a}& \cmark & \cmark &\cmark & $O(1)$ & $O(1)$&  \cite{orion,deepRDMA,shuffle} & Read\tnote{a} & No \\ 
                             & Indirect read & 2   & 1\tnote{a} & 1\tnote{a}   &(\cmark/\xmark)\tnote{a} & \cmark & \cmark &\xmark & $O(1)$ & $O(1)$&  \cite{nvmeof}  & Write\tnote{a}  & (No/Yes)\tnote{a}   \\  \hline
\multirow{3}{1.7cm}{Read Ring buffer} & Detached bell &$\geq4$\tnote{c}&0 &$\geq2$\tnote{c}& \xmark & \xmark & \cmark\tnote{d} &\cmark & $O(1)$ & $O(1)$& this work & Read & No   \\ 
                                    & Inlined bell  &$\geq4$\tnote{c}&0 &$\geq2$\tnote{c}& \xmark & \xmark & \cmark\tnote{d} &\cmark & $O(1)$ & $O(1)$& this work & Read & No   \\  
                                      & Notify receiver  & 3 & 1 & 1 &  (\cmark/\xmark)\tnote{a}& \xmark & \cmark & \cmark &  $O(1)$ & $O(1)$&  this work & Read\tnote{a} & No  \\ 
                                      \hline 
                                
\bottomrule
\end{tabular}
\begin{tablenotes}
\item[a] Depends on the employed second channel. 
\item[b] Requires appending extra data to a message.  \item[c] Pulling a new bell can take multiple requests.
 \item[d] Prefetching may incur additional copies.
\item \textbf{N/A}  - \ Not applicable.
\item \textbf{HRT}  - \ Half round trip.
\item \cmark ~- \ provides support
\item \xmark ~- \ does not support
\end{tablenotes}
\end{threeparttable}%
}
\caption{Performance and design features offered by uni-directional data channels for the sender (Send) and the receiver (Recv). Each protocol includes requirements for the transport protocol to implement them. }
\vspace{-0.5cm}
\label{tab:protocols}
\end{table*}

\section{Exchange Protocols}\label{sec:protocols}
We implemented~\cite{source} and analyzed twenty uni-directional data channel protocols that can be used as building blocks for any networked system.
All proposed protocols are fundamentally different in memory management (i.e., how memory is organized on the sender and the receiver to communicate data) or in network management (i.e., how the sender and the receiver interact with the RNIC to enable communication). We group channels into six categories/families depending on memory organization and the used underlying RDMA requests. For each channel, we indicate supported networking features and requirements to the underlying RDMA transport in Table~\ref{tab:protocols}.  
%
%Later in Section~\ref{sec:newsystems} we discuss how system developers can exploit this analysis to build right communication modules according to the needs of their systems. 

%\subsection{Implementation tricks}
\label{sec:tricks}
Before diving into data channels, we outline hardware and software tricks that can be used to enable some protocols. The former tricks require special RNIC capabilities, whereas the latter can be implemented on any modern operating system.

\textbf{Hardware tricks}. 
%Here, we show how  RNIC capabilities listed in Section~\ref{sec:rdma} can be used to implement and improve data channels. 
RNIC capabilities listed in Section~\ref{sec:rdma} are extensively used to improve performance of systems~\cite{anujRDMA,herd,corm,Dragojevic2017RDMART,zero-copy-sges}. We argue the inlining and on-demand paging capabilities do not fundamentally affect the memory and data layers of networked systems as they only reduce overhead on DMA engines or simplify memory registration.
In contrast, the other capabilities, even though they have been introduced for improving performance, may offer new memory management and networking features to networked systems.

%\emph{In contrast, the first three capabilities, even though they have been introduced for improving performance, may offer new memory management and networking features to networked systems.} Scatter-gather lists can be used to append data to immutable buffers, thereby avoiding the need to copy them to intermediate mutable buffers. Shared receive queues allow applications to dynamically share receive memory buffers between several RDMA connections. Finally, device memory offers additional ultra-fast low-latency memory regions, residing near the network, to applications. 
%We elaborate more on fundamental features and how they can be enabled with these capabilities in Section~\ref{sec:protocols}.  

Scatter-gather lists can be used to achieve true zero-copy in the case when a channel needs to append protocol-specific prefixes or suffixes to immutable data buffers. These extra data can include data length or special signaling bytes that are expected by the receiver. For example, scatter-gather lists have been used to implement a zero-copy serialization library~\cite{zero-copy-sges}. Shared receive queues can be used to share receive buffers across multiple RDMA connections, significantly decreasing memory usage for the N:1 communication. 
Device memory offers additional memory regions located closer to the network compared to the DRAM, thereby accelerating applications that often send recently received data~\cite{nicmem}.
%In addition, ATOMIC requests, which have limited throughput~\cite{anujRDMA}, can achieve significantly better performance with the device memory. 

\textbf{Software tricks}.
Existing RDMA protocols use virtual addresses in their RDMA requests to access memory. The use of virtual addresses allows systems to avoid fragmentation of physical memory at anything coarser than page granularity. RDMA protocols can take advantage of virtual addresses as operating systems do. Importantly,  physical memory segments can be mapped to several virtual addresses allowing to access the same physical memory via different memory addresses. 
MICA~\cite{mica} and FaRM~\cite{farm} use this trick to build a virtually circular RDMA-accessible buffer by mapping the virtual memory addresses right after the end of the buffer to the physical pages of the original buffer, thereby making the end of the buffer appear circular and contiguous in virtual space.  As a result, local and remote accesses to such circular buffer can be performed without range checking and RDMA writes near the end of the buffer do not cause splitting an RDMA request into two requests.  
%Previously, this trick was exploited to enable compaction in RDMA-enabled remote memory systems~\cite{corm}.

\subsection{Send/Recv protocols}
The Send/Recv protocol family, where a sender uses SEND requests to send data and a receiver pre-posts fixed size RECEIVE requests to get data, offers three different channels: \textit{Normal}, \textit{Shared}, and \textit{Bufferless}. These channels do not support variable messages as one SEND request consumes one RECEIVE request regardless of message sizes. 
As messages are received to pre-posted buffers controlled the receiver, the sender has no control over the exact destination of the message, making the protocols not zero-copy on receive.
%If the receiver posts a buffer smaller than the upcoming SEND requests, the network error will be generated. 

\textbf{Normal.}
The sender and the receiver use a private channel to communicate data, limiting the receiver to receive messages only from one sender, entailing large overhead for the N:1 communication pattern.  Senders with reliable transport often utilize network acknowledgments as an implicit data channel to confirm data reception. However, if the receiver does not promptly post RECEIVE requests, the reliable transport will cause delivery failure, which causes costly connection disconnects for InfiniBand and a request timeout for EFA and 1RMA. Therefore, systems often explicitly inform peers about the number of posted RECEIVE requests to prevent network failures. 

%\begin{figure}[t]
%\centering
%    \includegraphics[keepaspectratio,width=0.96\linewidth]{img/map.pdf}
%    \vspace{-0.5cm}
%    \caption{A buffer with double mapped memory. The starting virtual address $s$ is page aligned. }
 %   \vspace{-0.4cm}
%    \label{fig:map}
%\end{figure}

\textbf{Shared receive.}
Shared receive allows the receiver to share receive buffers between multiple senders.  If not enough buffers are pre-posted, the reliable transport of InfiniBand is subject to connection failures during network bursts.
Overall, shared receive has the same features as the normal version but helps to reduce memory usage for the N:1 communication.

\textbf{Bufferless.}
Bufferless data channel allows sending an integer using payload-less messages with IMM data. The receiver must still post RECEIVE requests to receive IMM data but they may have no buffers attached. This channel is often used to enable low-latency notifications without the need to register and manage memory.

\subsection{Write slot protocols}\label{sec:writeslot}
Write slot  protocols rely on writing messages to pre-allocate fixed size memory regions, called \emph{mailboxes} or \emph{slots}, using WRITE requests. The main property of this family is that the sender and receiver support zero-copy. As WRITES are silent, we outline all approaches to inform the receiver about incoming messages. 

\textbf{Inlined bell}. This method reserves a special memory location within each mailbox slot that indicates that a message is fully written. The sender writes a message into the mailbox so that it changes the state of the bell field. After processing, the receiver clears the bell to reuse the slot. In some cases, systems use remote memory as temporal remote storage that is not required to be notified about incoming data. For example, InfiniSwap~\cite{infiniswap} silently swaps out pages to remote memory to fetch them later when they are required. We assign such use-cases to the inlined bell category since they both logically manage memory as plain fixed size mailboxes.

\textbf{Detached bell}. This method stores the bell outside the mailbox slot. In this case, the sender writes data into the mailbox and then sends an additional request into the bell to signal the message reception. Unlike the inlined bell approach, this channel requires sending two RDMA messages that must come in-order as the signal should arrive after the message is fully received. 
%Note that some RNICs can offer user-defined memory layouts~\cite{umr} allowing a single WRITE to be scattered to two different memory locations. In this case, a detached bell becomes an inlined bell as only one WRITE is used. 
% nvme uses it

\textbf{IMM-based}. This method uses one WRITE\_WITH\_IMM request to signal the message arrival via a completion event at the receiver. The sender can encode into the IMM data the used mailbox. 

\textbf{Reserve slot}. The main disadvantage of the write protocols is that they have high memory usage, especially for the N:1 communication. The limitation comes from the fact that slots cannot be shared and the receiver allocates slots for each client individually. To address this issue, we can introduce a data path to lock and unlock slots for a specific sender. The sender can use any two-way protocol to reserve slots before using them and then later unlock them, thereby enabling buffer sharing between senders at the cost of extra round trips. The slot reservation can be implemented via bufferless sends to reduce memory usage. Note, depending on the implementation, one sender can take up all slots preventing other clients from sending messages.   

\subsection{Ring buffer protocols}\label{sec:rings}
Ring buffers rely on writing messages to a remote pre-allocated buffer with WRITE requests. The pre-allocated buffer forms a circular ring, allowing the sender to write messages one by one into the same region.
To enable efficient implementation of circular buffers, systems use the software trick that maps the virtual address after the ring buffer to the beginning of the buffer, allowing to access ring buffer if it was truly circular in memory (see Section~\ref{sec:tricks}). 
%As a result, reads and writes to the ring buffer can be implemented without handling writes beyond the buffer size. 
We distinguish the ring buffer algorithms by the way of informing the receiver about the incoming messages.

\textbf{Detached bell}. In this case, the detached bell stores the current head offset of the ring buffer. 
The sender writes a message into the ring buffer and then writes the new head into the bell. The RDMA messages need to come in-order as the bell should arrive after the message is written. To send messages of variable length, the sender needs to append the message length to each message. 

\textbf{IMM-based}. This algorithm uses WRITE\_WITH\_IMM to generate a completion event containing the message length at the receiver after writing the data to the ring buffer, enabling blocking receive.
%. The receiver needs to post empty RECEIVEs to get completion events, 

\begin{figure}[t]
\centering
    \includegraphics[keepaspectratio,width=1\linewidth]{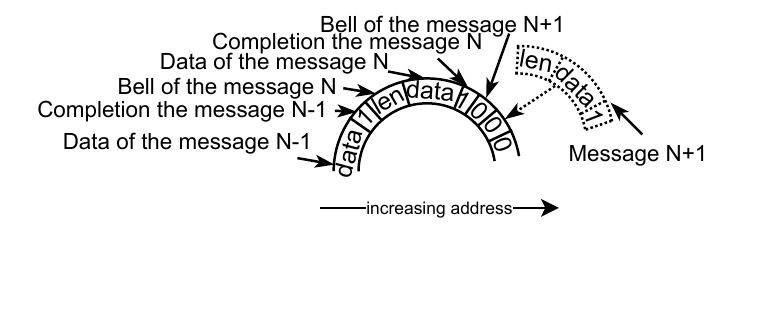}
    \vspace{-1.9cm}
    \caption{A ring buffer with the inlined bell with zeroing.}
    \label{fig:ring}
    \vspace{-0.5cm}
\end{figure}

\textbf{Inlined bell with zeroing}. This method appends the message length before the message as a bell.
In addition, it appends the value "one" to the end of each message to inform the receiver that the message is fully written (see Figure~\ref{fig:ring}). The sender writes the length, data and the completion with one WRITE request. The receiver polls the length field of the upcoming message of the circular buffer and once it is not zero it polls the inlined completion after the message until it is not zero. Once the message is processed, the receiver needs to clear the memory region occupied by the message to unset all possible bells and completions of future messages.

\begin{figure}[t]
\centering
    \includegraphics[keepaspectratio,width=1\linewidth]{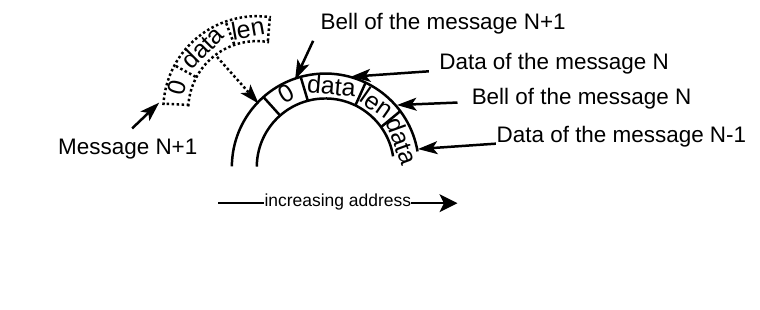}
    \vspace{-1.9cm}
    \caption{A ring buffer with the inlined bell without zeroing. }
    \label{fig:reverse}
    \vspace{-0.25cm}
\end{figure}

\textbf{Inlined bell without zeroing}. 
The main disadvantage of the previous algorithm is the need to clear memory after message processing. We found that it is possible to implement a ring buffer without zeroing and still use a single WRITE request. 
For that we flip the order of messages in the ring buffer:  instead of placing a new message after the last message, we place the new message before it in memory. Figure~\ref{fig:reverse} shows how writing a message works. Each message starts with the value "zero" and is followed by the message length. 
When a new message is placed, the value "zero" unsets the bell for an upcoming message and the length value sets the bell of the current message, which is at the end of the current message and at the beginning of the previous message.  
As a result, the sender zeros the bell of the upcoming message, and the receiver does not need to clear all memory. The protocol requires in-order byte and message delivery, as written bytes must be strongly ordered even across messages.% To the best of our knowledge, we are the first to introduce it. 

\subsection{Shared ring buffer protocols}\label{sec:sharedrings}
Ring buffer protocols suffer from high memory usage when a receiver needs to receive messages from many endpoints. To reduce the memory usage we propose shared ring buffer protocols that allow sharing one ring buffer between many writers. Each sender before writing a message needs to reserve a slot in the shared circular buffer using ATOMIC operations.
The sender atomically fetches the current head offset of the ring and adds the length of the message it wants to write. Then it writes the message to the fetched offset. 
Note that each writer needs to ensure that the receiver processed the previously written messages to that offset of the ring buffer. Thus, each writer sometimes needs  to read the tail of the remote ring, which indicates the processed offset.

\textbf{Inlined bell with zeroing}.
After fetching the head, each sender writes messages as in the corresponding ring buffer protocol.
Though senders can write messages in any order after fetching the head, the receiver can process them only in the offset order, since it only knows where the next message starts.  Therefore, a slow writer can block message processing.

\textbf{IMM-based}. It is an extension of the ring buffer with IMM, where the sender encodes the fetched head into IMM data, allowing the receiver to process messages in any order.  

\textbf{Impossibility of other protocols.}
The reverse buffer cannot be employed to receive messages without zeroing, as it is impossible to order WRITEs between different connections. Thus, one writer can accidentally unset the bell of the unprocessed message. 
We cannot have a detached bell protocol either, as the detached head at the receiver only indicates the intention of writers to write. Even with a second detached bell storing written bytes, clients could increment that bell in any order, falsely acknowledging incomplete messages from slow writers. 
If the writers use a two-directional channel to reserve slots instead of ATOMIC requests, then the protocol becomes the "write slot with reserve slots" method.

\subsection{Read slot protocols}
Read slot protocols rely on using READs to fetch data from predefined locations with zero-copy. This family of algorithms is not a good fit for message passing and therefore is usually used in combination with other channels that are used to inform receivers about readable memory regions. 
 
\textbf{Inlined bell}. This method inlines a bell within the mailbox slot that would indicate that the message is ready to be read. The receiver fetches the region and checks the bell. We also assign all algorithms that read a predefined region using one READ request into this category. For example, InfiniSwap~\cite{infiniswap} reads remotely swapped pages, and Derecho~\cite{derecho} fetches remote version tables. 

\textbf{Detached bell}.  
In this protocol, the sender writes to the detached bell a special value that can be read by a remote user with a READ. Once the receiver fetches the bell it can fetch the record.
The sender often encodes into the bell the address from which records have to be fetched by the receiver. Thus, we assign all pointer traversing algorithms to his category. For example, Wei et al.~\cite{xstore} and Ziegler et al.~\cite{tree-rdma} used READs to follow pointers in a remote tree structure.
%Some KVSs use this approach as well. In FARM-KVS, the overflow list on the table can be assigned to this algorithm. 

\textbf{Notify receiver}. This approach relies on actively informing the receiver about messages that could be fetched. The sender, for example, can use the bufferless send protocol for that. 

\textbf{Indirect read}. The key approach to having a flexible pull channel is to pre-allocate a receive buffer at the initiator and send its address to the sender. The sender can now use a write slot channel to write the data to the pre-allocated buffer. After the receiver receives the data, it can de-allocate the buffer and reuse memory for messages from other endpoints. As a result, the sender and the receiver can have constant memory usage regardless of the number of endpoints. 

\subsection{Read ring buffer protocols}\label{sec:readrings}
The previously discussed read slot channels are general methods for fetching data and they are not practical for proper messaging. To address this issue we propose circular buffers for senders that offer an interface for sending messages to multiple remote readers. In this family, the sender writes messages \emph{locally} to an RDMA-readable circular buffer that is read by remote clients with READ requests. 

\begin{figure}[t]
\centering
    \includegraphics[keepaspectratio,width=0.9\linewidth]{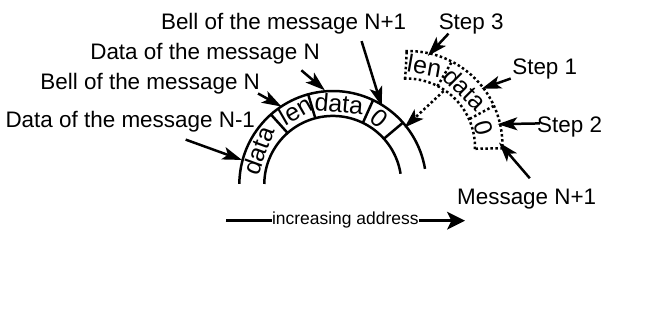}
    \vspace{-1.5cm}
    \caption{A read ring buffer with the inlined bell. }
    \label{fig:readring}
    \vspace{-0.5cm}
\end{figure}

\textbf{Inlined bell}. The sender writes a message in the circular buffer with inlined bell in three steps  (see Figure~\ref{fig:readring}): first, it locally copies the message into the buffer, then it clears the bell after the message, and then it writes the length into the bell before the message. The readers fetch the length and then read the full message. It is crucial that these local writes are strongly ordered by the CPU (e.g., using memory fences).  Each receiver can safely read more bytes than the size of the length field to prefetch more data. 

\textbf{Detached bell}. This method stores the head of the circular buffer in the detached bell. Each reader first fetches the head value and then they fetch new messages if the head value advanced. The downside of this approach is readers do not know how many messages they fetch unless the protocol is used for fixed size messages. 

\textbf{Notify Receiver}. This method actively informs receivers (e.g., via bufferless send) about new messages in the circular buffer . 

\section{Systems design guidelines}\label{sec:newsystems}
Table~\ref{tab:protocols} shows that protocols offer different features and often support only certain RDMA networks. 
Developers can use our study to find channels according to their system requirements and capabilities of targeted networks.
They can choose protocols that run on all interconnects, thereby making their implementation universal. They can also help to find protocols that address primary needs such as low memory usages, scalability, or low latency.

Besides facilitating correct design of new systems, our study can be used
to identify systems whose communication channels contradict
the core requirements of primary workloads.
For example, RDMA-enabled swapping systems aim to swap out 4 KiB pages to remote daemons with low-latency, low CPU cost, and efficient memory management at the receiver, which can be satisfied with the "Shared Send/Recv" protocol. However, all existing RDMA-enabled swapping systems~\cite{infiniswap,far} employ "write slot with inlined bell", thereby wasting a lot of memory at the remote swap daemon.

\section{Evaluation}
To demonstrate the correctness of the studied uni-directional channels we have implemented them and measured their performance under various workloads and deployment settings. 
With our evaluation we aim to answer the following questions:
%that we find particularly interesting:

\begin{itemize}[noitemsep,topsep=0.00cm,leftmargin=0.3cm]
\item Does true zero-copy requirement allow networked systems to improve their performance?
\item Do our performance metrics (HRT and \# Requests) reflect the empirical performance of the protocols?
%\item What is the empirical contribution of scatter-gather lists on the performance of ring protocols?
\item Does our "ring inlined bell w/o zeroing" outperform other ring implementations?  %+
\item Can the device memory capability of RNICs improve the performance of shared ring protocols? %+
\item What performance can be achieved by shared ring protocols for the N:1 communication and by read ring protocols for the 1:N communication? %+
\end{itemize}

\textbf{Experimental setup.}
All experiments were performed on two machines interconnected by 100 Gb/sec Nvidia  ConnectX5 NICs. Machines communicate with each other via a switch using RoCEv2 protocol.
Each machine is equipped with AMD EPYC 7742 @ 2.25GHz and 256 GiB of DRAM.

\textbf{Implementation details.} %
All protocols and systems are implemented
%\footnote{\href{https://www.dropbox.com/s/ps3fkca7hjf2it6/RDMA-and-systems.tar.gz}{www.dropbox.com/s/ps3fkca7hjf2it6/RDMA-and-systems.tar.gz}}
in C++ and depend on the following libraries: \textit{libibverbs}, an implementation of the IB verbs, and \textit{librdmacm}, an implementation of the RDMA connection manager. 
All channels are implemented as single-threaded message interfaces that expose the following API:

\begin{figure}[H]
\vspace{-0.35cm}
 \begin{minted}
[
frame=lines,
framesep=0.1mm,
baselinestretch=0.1,
fontsize=\footnotesize,
]
{cpp}
// Sender API
// Send a region to the receiver. The send will be
// zero-copy if the protocol is zero-copy on send.
uint64_t SendRegion(Region *r); 
// To test the completion of the send request
bool TestSendRequest(uint64_t handler); 




// Receiver API without zero-copy on receive.
// Receive the next message.
Region* ReceiveRegion(); 
// Acknowledge the region processing to reuse it.
bool FreeReceiveRegion(Region *r); 




// Receiver API with zero-copy on receive.
// Receive the next option to receive a message.
RegionOption* CanReceiveRegion(); 
// Initiate zero-copy message reception.
uint64_t ReceiveRegion(RegionOption *o, Region *r);
// To test the completion of the receive.
bool TestReceiveRequest(uint64_t handler); 
\end{minted}
\vspace{-0.45cm}
\end{figure}

As mentioned earlier in Section~\ref{sec:comprot}, a network module should include at least two uni-directional channels in opposite directions to build a proper datapath with reliable data delivery: a receiver needs to acknowledge message reception and message processing. For example, senders of all ring-based protocols need acknowledgements to prevent overwriting unprocessed messages with new messages in the remote ring. 
Networked systems can freely combine the studied protocols to build network modules according to their needs.
In our tests, we implement the following bi-directional channels:

\begin{itemize}[noitemsep,topsep=0.00cm,leftmargin=0.3cm]
\item All point-to-point channels are duplicated to be used in both directions, making each endpoint both the sender and the receiver of a particular point-to-point channel.
\item Each receiver of read ring protocols is equipped with the "Write slot inlined bell" protocol to write its processed tail to the sender, so the sender can advance the tail of the shared ring by finding the smallest tail among its readers. We chose that protocol as it preserves the network passiveness (see Section~\ref{sec:systems}) of the sender.
\item Each sender of shared ring protocols uses "Read slot inlined bell" to occasionally fetch the tail of the shared buffer. With the information about the tail, each sender can avoid overwriting unprocessed messages.  We chose that protocol to preserve the network passiveness of the receiver.
\end{itemize}

%\textbf{Progress tracking.}
%The sender, which writes locally message, should also track the progress of readers to prevent the overwriting of not sent messages. For that we propose to use Direct Write Magic byte algorithm to send progress of the receivers to the sender. Each receiver periodically writes its head to the memory of the sender. The sender finds minimum "watermark" over the array of heads and updates its local tail, which indicates the pointer the oldest not replicated message.

\subsection{Performance microbenchmarks} 
\textbf{Zero-copy requirement.}
To motivate the importance of zero-copy communication for networked systems we measure the performance of the two-directional protocol based on the send/recv normal channels with and without the need to copy received messages. 
In the experiment, a client sends a message to a system via the channel, and the system after processing the request sends a response back.
The response size is the same as the request size. 
In the "with copy" setting, the system needs to copy the content of the received message to log-structured in-memory storage with the capacity of 8 GiB.

\begin{figure}[t]
\centering
    \includegraphics[keepaspectratio,width=0.9\linewidth]{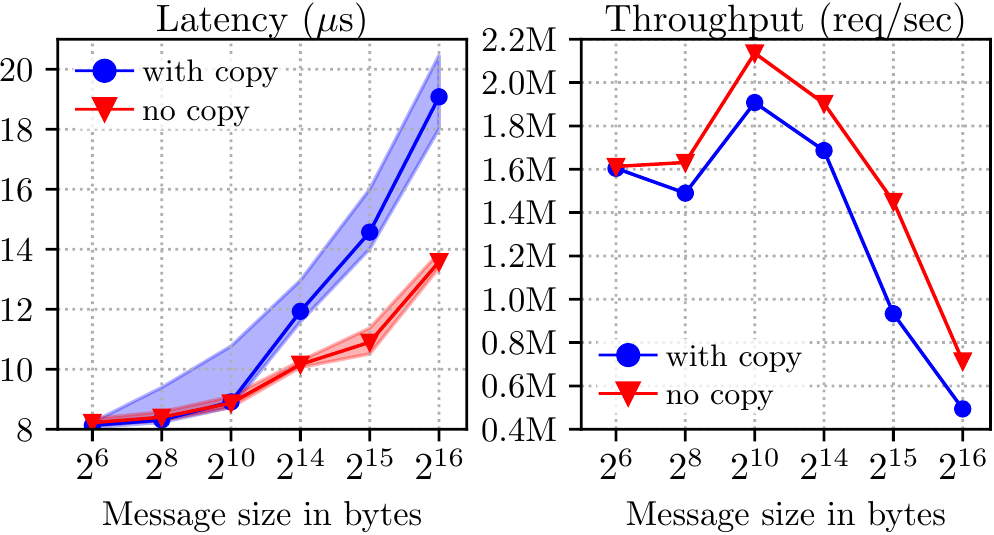}
    \vspace{-0.4cm}
    \caption{The round-trip latency and throughput of the send/recv channel with and without additional memory copy.}
    \vspace{-0.45cm}
    \label{fig:copy}
\end{figure}

 \begin{figure}[t]
\centering
    \includegraphics[keepaspectratio,width=0.9\linewidth]{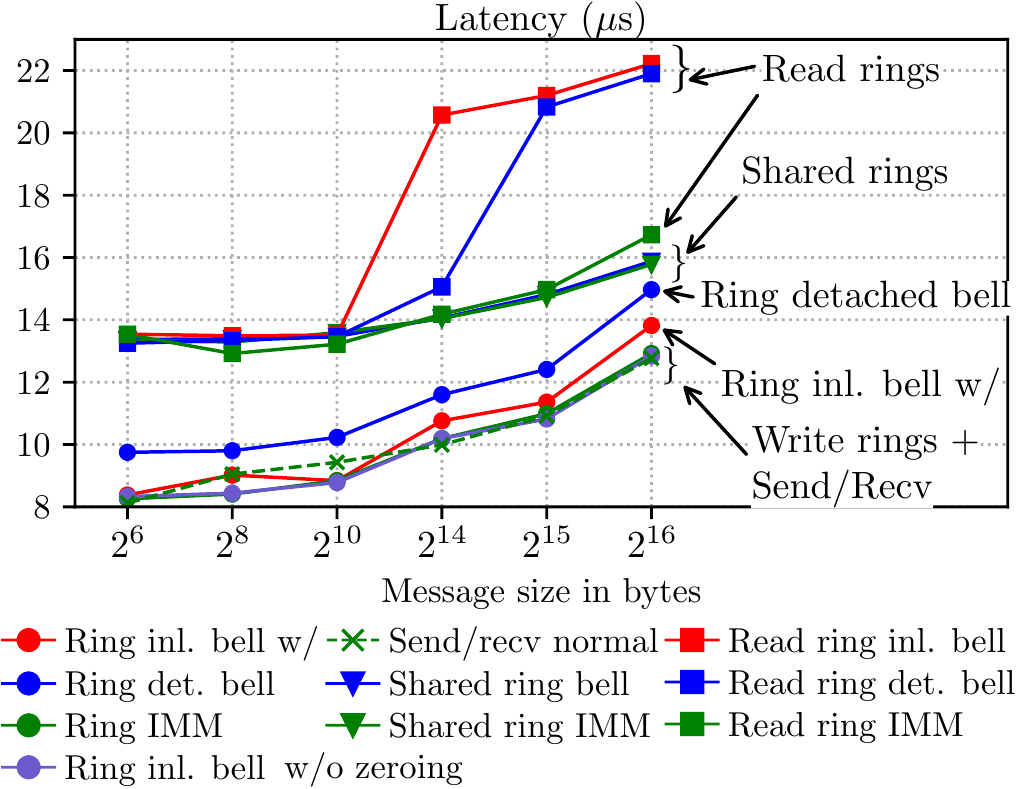}
    \vspace{-0.35cm}
    \caption{Median round-trip latency of various channels.}
    \vspace{-0.1cm}
    \label{fig:lat}
\end{figure}

Figure~\ref{fig:copy} shows that the copy requirement caused the client to experience higher request latencies with higher deviation (the 90\% confidence interval is depicted). The difference becomes more pronounced for large messages. In addition, the average throughput was by 0.2M req/sec higher for the zero-copy system, showing the advantage of the zero-copy.

\textbf{Performance metrics.}
To show that the latency of the data channels can be expressed as the number of half-round trips, we measure the round-trip latency of various channels with different performance characteristics. 
For the point-to-point channels, the latency includes the server's reply message that has the same size as the original message. For shared and read rings, the receivers do not send the whole message back and only acknowledge it by sharing the tail of the ring. 

Figure~\ref{fig:lat} shows that shared and read rings have the highest median latency as they require multiple HRTs to send data. All benchmarked channels requiring one HRT have approximately the same latency. However, the "ring inlined bell with zeroing" has a bit higher latency for large messages as the receiver needs to zero processed messages before sending a reply. The data also reveals that the "ring detached bell" channel has higher latency than other rings as it requires two work requests per message.

\begin{figure}[t]
\centering
    \includegraphics[keepaspectratio,width=0.85\linewidth]{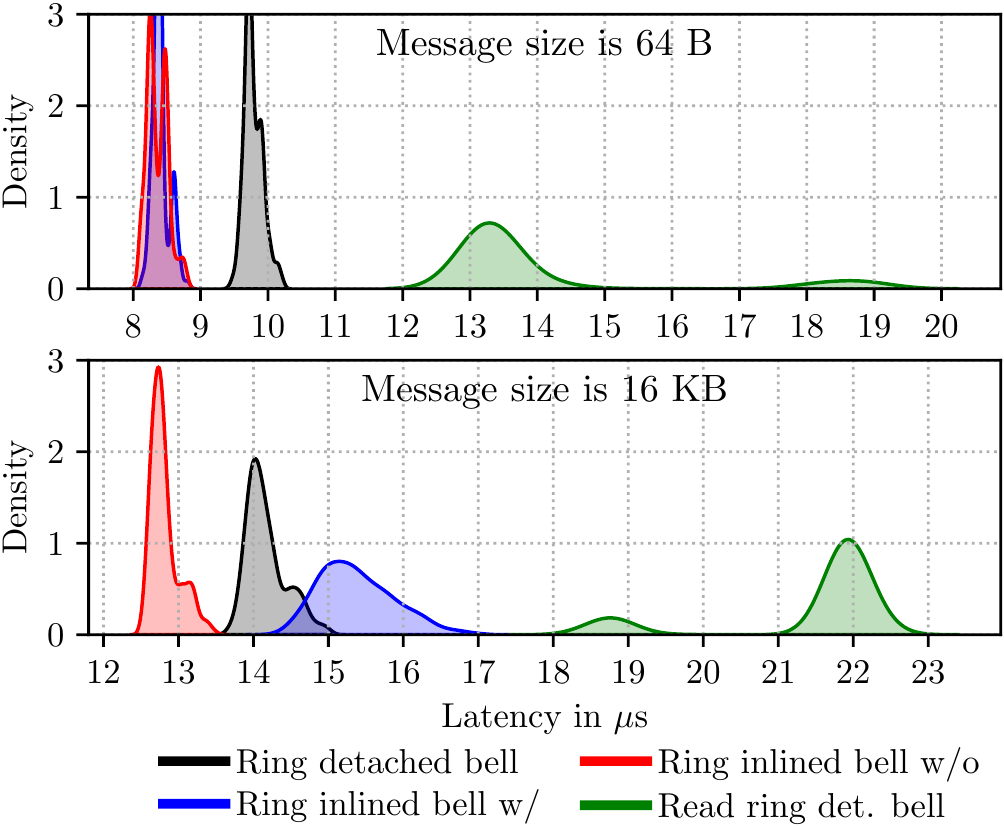}
    \vspace{-0.35cm}
    \caption{Probability distribution of the latencies.}
    \label{fig:prob}
    \vspace{-0.3cm}
\end{figure}

Interestingly, passive read rings have a dramatic drop in performance for large messages in Figure~\ref{fig:lat}, which reports the median latency.
Figure~\ref{fig:prob} helps to understand this artifact by providing empirical probability distribution of the latencies. The data reveals that the read ring channel follows a bimodal distribution. The first peak depicts latencies when the receiver fetched the new bell with one READ request. The second one is when the fetching the new bell took two READ requests. 
As the time required to write a message into the ring and update the bell depends on the message size, we can see the receiver was mainly in-sync with the sender for the small message. For the large message, however, the receiver was less fortunate and needed the second request.

%\textbf{Scatter-gather lists.} Some algorithms from Table~\ref{tab:protocols} required appending metadata to data buffers, which is not possible for immutable buffers. In the case of the immutable buffers, the sender would need to copy the buffer to the mutable space, incurring an additional copy. Figure~\ref{fig:sge} shows that the scatter-gather capability of the tested RNIC does not significantly change the latency of the ring channels and allows preserving the zero-copy on send feature.

% \begin{figure}[t]
%\centering
 %   \includegraphics[keepaspectratio,width=0.9\linewidth]{img/sge-crop.pdf}
  %  \vspace{-0.35cm}
   % \caption{The effect of scatter-gather entries (sge) on the round-trip latency of ring-based protocols.}
%    \vspace{-0.3cm}
 %   \label{fig:sge}
%\end{figure}

\subsection{Point-to-point channels} 
In Section~\ref{sec:rings}, we have presented a new approach for delivering data to a ring buffer that does not require zeroing memory when the inlined bell is used. To evaluate the effectiveness of the proposed method we measure the throughput of all ring-based channels and the send/recv normal channel.

 \begin{figure}[t]
\centering
    \includegraphics[keepaspectratio,width=0.9\linewidth]{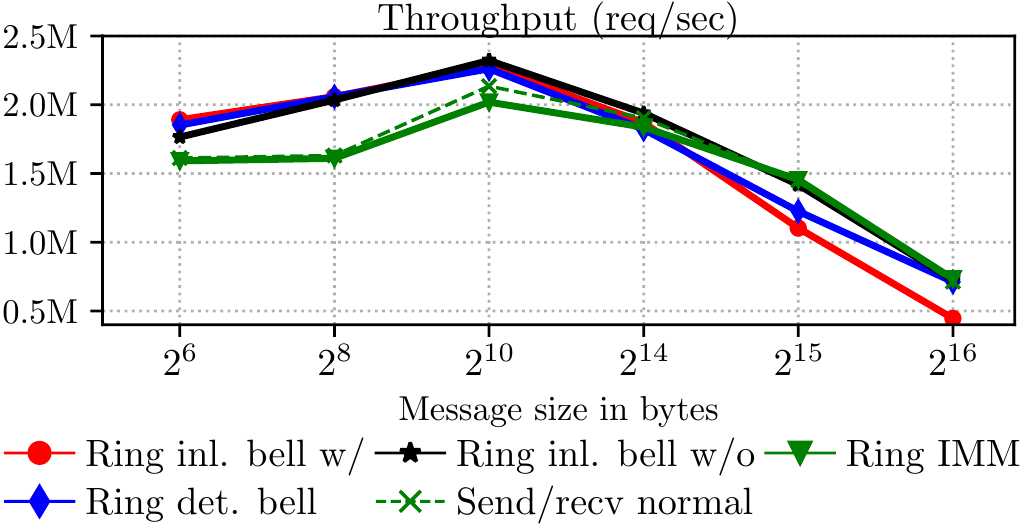}
    \vspace{-0.35cm}
    \caption{Throughput of ring-based and send/recv protocols.}
    \label{fig:bw}
    \vspace{-0.15cm}
\end{figure}

Figure~\ref{fig:bw} reports the average throughput of the discussed channels measured by a client of a system. The client could have multiple outstanding requests and the system was empowered to batch responses to ensure that the measured throughput is bottlenecked by the client to the server channel. 
The lowest throughput for small messages has been observed for methods requiring the receiver to post  RECEIVE requests, which limited the speed of the protocol.
For large messages, however, they achieve maximum performance.

Our inlined bell algorithm without zeroing outperforms the counterpart with zeroing, as it does not require the CPU to clear processed messages. Therefore, we conclude that our algorithm should replace the existing inlined bell method as they offer the same features to the networked systems. However, if the RDMA network does not offer in-order message delivery, then our algorithm cannot be used.

Due to the pipelining effect between outstanding requests, the detached bell method performed as the inlined bell method without zeroing for some sizes. Nonetheless, the second WRITE request always reduces the link utilization.

 \begin{figure}[t]
\centering
    \includegraphics[keepaspectratio,width=0.9\linewidth]{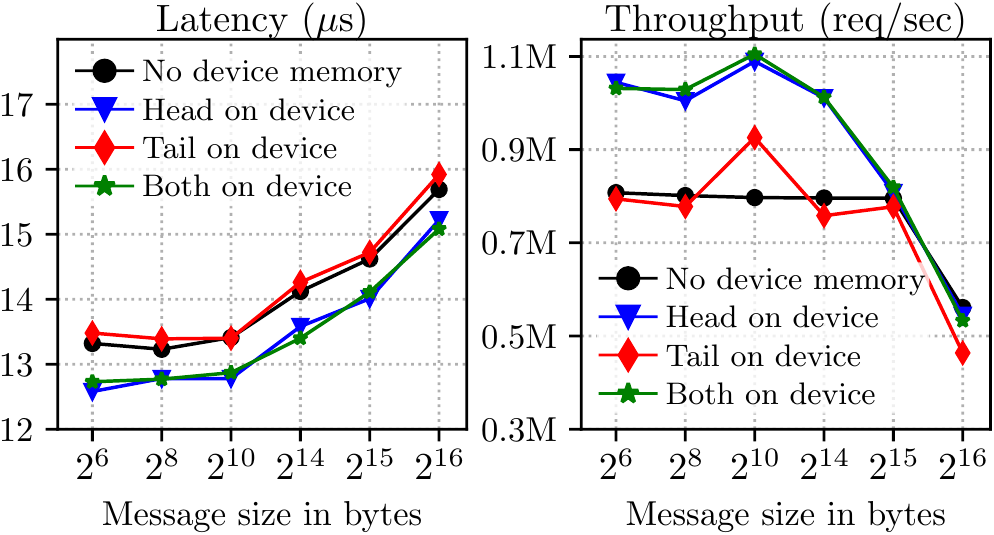}
    \vspace{-0.35cm}
    \caption{The effect of the device memory on performance of the shared ring with detached bell channel.}
    \label{fig:device}
    \vspace{-0.45cm}
\end{figure}

 \begin{figure}[t]
\centering
    \includegraphics[keepaspectratio,width=0.9\linewidth]{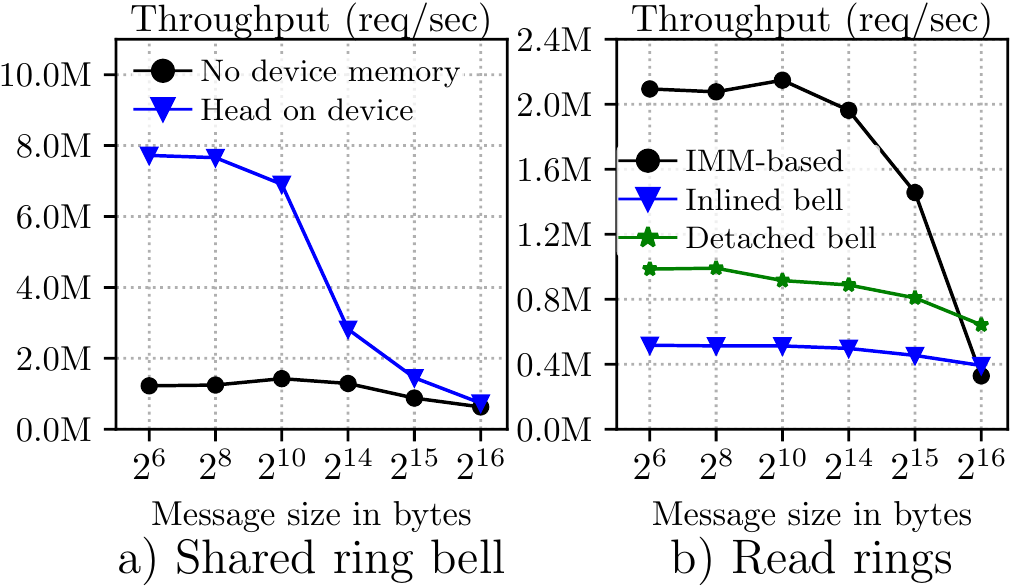}
    \vspace{-0.35cm}
    \caption{Cumulative throughput of shared and read rings under the load from multiple clients.}
    \label{fig:shared}
    \vspace{-0.3cm}
\end{figure}

\subsection{Shared and read ring channels} 
\textbf{Device memory.} 
The device memory extends the memory of networked systems with a small memory region that is closer to the network than the main memory. As the device memory is located at the RNIC, RDMA requests to it can be processed without PCIe involvement. To measure the role of the device memory on channels, we measured the performance of the shared ring inlined bell method with various deployment settings depending on the memory location of the head and the tail of the shared buffer. As a reminder, each sender issues an ATOMIC request (fetch and add) against the head to reserve a slot in the buffer, and occasionally a READ request against the tail to monitor the progress of the receiver.

Figure~\ref{fig:device} reports performance observed by a client of a system that used the shared ring with the inlined bell algorithm. As fetching the head is on the critical path, moving the head to the device reduced the latency, whereas the tail location had no effect. The difference between "no device memory" and "head on device" is equal to the round trip latency of the PCIe bus. The throughput also is increased for the cases when the head is in the device memory. However, the increase was not dramatic as we allowed the sender to have at most 16 outstanding messages. 

Figure~\ref{fig:shared}a shows the cumulative throughput of the shared ring with the inlined bell under the load of 8 clients, where each client could have at most 16 outstanding requests. The throughput is measured by the receiver. 
The plot shows that the performance of the channel has been bottlenecked by the performance of ATOMIC requests against the main memory. When the head has been moved to the device, the receiver experienced a more than 4x improvement in throughput,
showing that the device memory can be efficiently used to speed up synchronization between clients with ATOMICs.

 \begin{figure}[t]
\centering
    \includegraphics[keepaspectratio,width=0.9\linewidth]{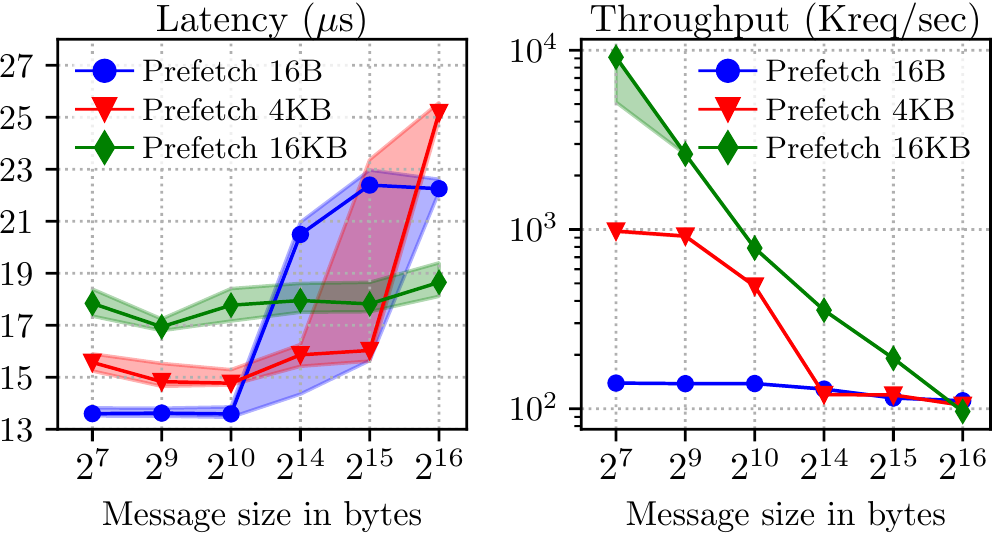}
    \vspace{-0.35cm}
    \caption{The effect of the prefetching on performance of the read ring with inlined bell channel.}
    \vspace{-0.3cm}
    \label{fig:prefetch}
\end{figure}

\textbf{Broadcast.} 
Read rings allow the sender to passively replicate data to multiple nodes without the need to work with RDMA requests.  
 We compare the performance of proposed algorithms under a workload of four clients that receive messages from the server by reading them from the read ring. 

Figure~\ref{fig:shared}b shows the cumulative throughput of the read rings under the load of 4 clients. The IMM-based algorithm has the highest throughput as its clients do not waste the network on fetching the bell, and only issue READs to fetch data once they receive a bufferless message containing the current head position. The detached bell performs better than the inlined one, as the detached bell allowed the clients to prefetch multiple records (by looking at the current head). The inlined bell was allowed to fetch only one record at one READ (as it was polling on a bell of each message).

\textbf{Prefetching.} Figure~\ref{fig:prefetch} shows the performance of the read ring with inlined bell when a reader can optimistically read the data after the current bell. Prefetching improves the throughput of the receiver as it can read multiple records, but has a negative impact on the latency of small records. For large records, we see the same latency artifact as in Figure~\ref{fig:prob}.

\section{Future RDMA networks}\label{sec:rdmafuture}
We believe that algorithms listed in Table~\ref{tab:protocols} will persist even with the next generations of RDMA products.
To show that we below discuss experimental and future RDMA networks known to us.
%We argue below that these new RDMA networks will only slightly transform proposed algorithms. 
% 

The \textit{multi-message receive} capability~\cite{multireceive,portals4} allows a RECEIVE request to be reused for multiple messages, where each message atomically takes a part of the receive buffer. Fundamentally, it will enable Send/Recv algorithms to support variable size messages.

The \textit{tag-matching} capability~\cite{tagmatching} would allow a receiver to dynamically choose the target location of SEND requests.
Fundamentally, it may facilitate zero-copy receiving, turning Send/Recv algorithms to the write slot with IMM protocol. Note that the shared receive buffers with tag-matching would reduce memory need to $O(1)$.

%\textbf{Optimized DMA gather transfers.}
Existing RNICs issue one DMA read per entry in the gather list of a send request, degrading PCIe performance for highly fragmented transfers.
D-RDMA~\cite{d-rdma} proposes to allow RNICs to freely generate any DMA transfer strategy that brings data 
from DRAM. Thefore, the RNIC can optimize its DMA schedule to reduce the number of requests as well as the amount of transferred data. 
%To enable optimizations across several requests, D-RDMA allows users to batch send requests.
This proposal offers improved performance for requests with good spatial locality (e.g., sending the same buffer to multiple nodes). Fundamentally, it does not change the system design and only reduces PCIe load.

%\textbf{Scattered one-sided requests.}
Some RNICs can offer \textit{user-defined memory layouts}~\cite{umr} allowing a WRITE request to be scattered to multiple different memory locations. In this case, algorithms that required multiple ordered WRITEs would require only one byte-ordered WRITE.

PRISM~\cite{prism} introduces a series of changes to RNICs to empower applications to perform complex read and write accesses with a single network round trip. For example, PRISM proposes an indirect read request that performs a READ in two steps: read a virtual address from the remote buffer and then read  data from that address to the initiator. Fundamentally, proposed extensions reduce the number of network trips but still require the same number of PCIe requests as all intermediate steps are synchronized at the remote RNIC, thereby partially reducing the latency of complex access patterns.

Multiple research papers~\cite{spin,strom} propose to extend RDMA requests with general-purpose functions that could perform complex memory accesses from the RNIC. Often such RNICs are called smartNICs as they can execute complex instructions during packet processing. Similar to PRISM, such proposals only reduce the network latency and not PCIe latency. Nonetheless, such smartNICs can partially offload complex protocols, proposing completely new data and network management opportunities as well as design challenges. %SmartNIC-enabled systems are outside of the scope of this work.

\section{Conclusions}
The choice of RDMA-based communication channels has a great impact on the design of networked systems and the requirements for the utilized RDMA networks.
We have analyzed and categorized data communication algorithms, helping system developers choose them according to their needs. We have introduced new communication algorithms and discussed the effect of future RDMA products on the studied algorithms, ensuring the comprehensiveness of the study.
 
{\bibliographystyle{plain}
\bibliography{references}}

\end{document}